# Temperature-Dependent Discovery of BCC Refractory Multi-principal Element Alloys: Integrating Deep Learning and CALPHAD Calculations

Ali K. Shargh[1*], Christopher D. Stiles[1,2], Jaafar A. El-Awady[1*]

[1] Department of Mechanical Engineering, Johns Hopkins University, Baltimore, Maryland 21218, United States

[2] Research and Exploratory Development Department, Johns Hopkins University Applied Physics Laboratory, Laurel, Maryland 20723, United States

# Highlights

- Our DL model preserves CALPHAD fidelity while significantly speeds up predictions
- The model predicts phase fractions for up to eight phases across temperatures
- Ti–Fe–Al–V–Ni–Nb–Zr space screened for stable single-phase BCC alloys
- Ni suppresses BCC phase while Fe and Al stabilize it at elevated temperatures
- Our interpretable equation enables fast screening of single-phase BCC RMPEAs

[*] Corresponding authors:
Email addresses: ashargh1@jhu.edu (A. K. Shargh), jelawady@jhu.edu (J. A. El-Awady)

# Abstract

Single-phase body-centered cubic (BCC) refractory multi-principal element alloys (RMPEAs) offer potential for developing alloys with exceptional strength. However, the compositional design space is immense. Exhaustively mapping this space with conventional CALculation of PHAse Diagrams (CALPHAD) is impractical because database coverage and run times scale poorly with millions of candidate chemistries. To address this, we train a deep-learning surrogate on CALPHAD outputs that preserves the thermodynamic fidelity while accelerating temperature-dependent phase-fraction predictions of RMPEA phases. The model achieves high accuracy in predicting phase fractions for up to eight distinct phases across different temperatures and offers a speedup of two orders of magnitude compared to CALPHAD. Using this model, we screen the Ti, Fe, Al, V, Ni, Nb and Zr elemental space for potentially stable single-phase BCC alloys at different annealing temperatures and extract design insights to guide the synthesis of new BCC RMPEAs in experiments. Finally, we develop an analytical model that enables rapid, interpretable identification of single-phase BCC RMPEAs.

# Keywords

Machine learning, Phase prediction, BCC alloys, CALPHAD, Surrogate modeling, Multi-principal element alloys, High entropy alloys

# 1. Introduction

Refractory multi-principal element alloys (RMPEAs) have attracted significant attention due to their exceptional properties, for example, in ultra-high-temperature aerospace components [1,2]. These alloys exhibit high strength, excellent ductility, high electrical resistivity, and low thermal expansion coefficients [3–7]. Unlike conventional alloys, which typically contain one primary element with minor additions of others, RMPEAs are composed of multiple principal elements, each with concentrations ranging from 5 to 35 atomic percent (at. %) [3,8–10]. The complex interplay of deformation mechanisms that are chemistry and phase-dependent directly control the properties of RMPEAs. Therefore, predicting the phases that will form in each composition is a crucial first step in alloy design because phase constitution directly controls properties. For example, the face-centered cubic (FCC) structure in an RMPEA improves ductility, whereas the body-centered cubic (BCC) structure enhances strength [11–15]. Additionally, intermetallic Laves phases enhance wear resistance [16], sigma ($\sigma$) phases degrade corrosion resistance [17–19], and amorphous phases retard the interdiffusion of atoms at lower temperatures [20].

Mapping the phase space of RMPEAs experimentally is daunting. In the Ti, Fe, Al, V, Ni, Nb, and Zr space alone, the combinatorial design space contains roughly 12 million distinct compositions (assuming alloys with 3 to 5 elements, and each element varies from 5 at. % to 35 at. % in 1 at. % steps). Even state-of-the-art high-throughput synthesis may examine only tens of thousands of samples annually, leaving orders of magnitude unexplored. Exhaustively mapping this space with conventional CALculation of PHAse Diagrams (CALPHAD) is also impractical because database coverage and run-times scale poorly with millions of candidate chemistries unless one has access to extensive parallel computation infrastructure.

To address these limitations, researchers have increasingly turned to machine learning (ML) for material optimization and design [21–23]. ML strategies for RMPEAs typically involve creating surrogate models that can classify or predict phase types using input features, such as elemental compositions, atomic size differences, and valence electron concentrations. Numerous studies have demonstrated the efficacy of ML in predicting two or three types of phases in RMPEAs [24–30]. For example, Noor et al.[24] trained a multi-layer perceptron (MLP) neural network using a dataset of 240 experimental RMPEAs. Their model predicted amorphous, solid solution, and intermetallic phases in their validation dataset with an accuracy exceeding 80%. More recently, Chen et al. [25] developed a framework that combined four ML models with an improved Dempster-Shafer evidence theory. Using a dataset of 761 experimental RMPEAs, supplemented by 400 virtual samples generated from a predictive conditional generative adversarial network (CGAN), their model achieved an accuracy of 94.78% predicting amorphous, solid solution (SS), intermetallic (IM), and SS+IM phases. Importantly, the annealing temperature is not explicitly accounted for in these studies, even though the phase stability of RMPEAs is highly temperature-dependent [31].

Developing models that accurately classify a larger distribution of RMPEA phases at different temperatures requires larger, more comprehensive training datasets. To generate such data, several recent studies have leveraged CALPHAD to simulate phase equilibria at selected temperatures [32,33]. Deep learning (DL) models, trained on datasets generated using CALPHAD, can be an efficient surrogate for CALPHAD calculations, enabling rapid and accurate phase predictions while significantly reducing computational costs.

To quantify a broad range of possible phase combinations, in our previous study [33] we trained a multi-layer perceptron (MLP) neural network using a dataset of 50,000 RMPEAs within the elemental system of Ti, Fe, Al, V, Ni, Nb, and Zr. The model used 34 input features and was trained on a dataset labeled with CALPHAD calculations. These 34 features, identified as essential for accurately predicting RMPEA phases, were selected from a comprehensive list of 50 features commonly used in the community. Our model demonstrated over 90% accuracy in classifying different combinations of possible phases at 850K, namely FCC, BCC, hexagonal close-packed (HCP), Ordered BCC (B2), Laves (C14, C15, and C36), Heusler, Sigma, and Liquid. Nevertheless, that framework and similar efforts [32,33] are restricted to a single annealing temperature. Our current study introduces a more comprehensive framework to predict RMPEA phase fractions across a temperature range of 850 K to 1891 K for 3-, 4-, and 5-element alloys within the Ti, Fe, Al, V, Ni, Nb, and Zr elemental space. The framework can regress the fractions of various possible phase combinations, including FCC, BCC, HCP, B2 (i.e., ordered BCC), Laves (C14, C15, and C36), Heusler, Sigma, and Liquid phases. It is worth noting that the focus on Ti, Fe, Al, V, Ni, Nb, and Zr in this study is due to the rising interest in these elements as key components for RMPEAs for high-temperature applications. These elements have a unique combination of properties, including high melting points, the ability to form stable solid solutions, and a wide range of atomic sizes and electronegativities, making them ideal candidates for high-performance applications in extreme environments.

Our trained surrogate DL model is then used to search over the same element space to identify composition-temperature domains where a single-phase BCC structure (i.e., the phase that maximizes high-temperature strength) is predicted. The resulting maps yield design strategies that can guide the experimental synthesis of novel high-strength RMPEAs [34–39]. Moreover, these maps support our objectives at the Center on Artificial Intelligence for Materials in Extreme Environments (CAIMEE) at Johns Hopkins University, enabling autonomous high-throughput design of novel BCC RMPEAs for high-temperature applications.

While the DL model can provide high accuracy, as will be discussed, its nonlinear nature often obscures physical insight. Therefore, we also derive a closed-form empirical criterion fitted to the surrogate predictions to provide an interpretable, high-accuracy analytical screening model for single-phase BCC stability. This study extends our earlier work [33] in four important ways. First, it incorporates temperature as an input variable, enabling phase fraction prediction across a wide range of annealing conditions. Second, it moves beyond simple phase classification to quantitatively predict the phase fractions of up to eight phases in RMPEAs. Third, we derive a closed-form mathematical equation for identifying stable single-phase BCC regions, offering both interpretability and utility for experimental alloy design. Fourth, we use the ML model to explore the composition-temperature design space for single-phase BCC formation, thus providing actionable design guidance for experimental synthesis.

This paper is structured to systematically address the objectives outlined above. The processes of database generation, data labeling, and the architecture of our deep learning framework are discussed in the methods section. A comprehensive analysis of our framework's performance focusing on three key areas: (1) validation of the model's accuracy; (2) exploration of the design space for single-phase BCC RMPEAs; and (3) development and evaluation of our mathematical model for identifying single-phase BCC RMPEAs are then discussed in the results section. Finally, the conclusion section synthesizes our key findings and discusses their implications for the field of RMPEA design.

# 2. Methods

Our DL framework's development starts with preparing a comprehensive dataset that is labeled with expected RMPEA phases as well as relevant predictive features. Within the seven elements of interest (Ti, Fe, Al, V, Ni, Nb, and Zr), a random sampling approach is employed to generate 50,000 unique compositions. The sampling strategy ensures a diverse representation of potential RMPEAs within these compositional bounds. Each composition consists of 3-5 elements, with individual elemental fractions ranging from 5% to 35% with 1% increments. The total number of possible RMPEAs in this design space is approximately 12 million. Each sample is then labeled with 51 features, as defined in the following section and the Supplementary Materials. These features represent various physicochemical properties crucial for phase prediction. To account for temperature-dependent phase stability, we calculate the expected phases for each sample at seven different temperatures using the Python package Thermo-Cal [40]. The latest version of the thermodynamic database (i.e., TCHEA6) is utilized for our CALPHAD calculations.

The temperature range for each sample extends from 850 K to 0.8 $T_{\mathrm{m}}$, where $T_{\mathrm{m}}$ is the estimated melting temperature and it is calculated using the mean mixture rule. Seven temperatures evenly spaced within this range are chosen for each RMPEA. This results in seven different temperature values (excluding 850 K) for each sample since each has a different $T_{\mathrm{m}}$. As mentioned later, this also implies that the training process spans a broad range of temperatures rather than being limited to the same fixed set of temperature points across all samples. Our study focuses on the following phases: FCC, BCC, HCP, B2 (i.e., ordered BCC), Laves (C14, C15, and C36), Heusler, Sigma, and Liquid. When multiple phases share the same crystal structure but differ in composition (e.g., due to spinodal decomposition), we label them collectively based on structure. For example, samples containing Laves#1+ Laves#2 phases are labeled Laves.

The resulting dataset thus has 350,000 cases (50,000 compositions × 7 temperatures). Of those, Thermo-Calc could not converge on a solution for 1576 compositions, which are therefore removed from the final dataset.

Following data generation, a rigorous data engineering process is implemented to refine our dataset and feature set. This process is based on two main steps (see full details in our previous study [33]):

1) Feature correlation analysis: We eliminate one feature from each pair of strongly correlated features (|Pearson correlation coefficient| > 0.9). This step helps to reduce redundancy and potential multicollinearity in our model inputs. The Pearson correlation coefficient matrix for the 51 input features is shown in Supplementary Figure S2.
2) Feature distribution optimization: All features are normalized to the range (0,1) using min-max scaling. We then analyze feature histograms (bin size = 0.1) and exclude data points from bins with intensity < 20 material data points. This step ensures more uniform feature distributions across the dataset, which can improve model training and generalizability.

This process results in a unique set of 344,583 compositions and a reduced input feature set of 35. These input features include mixing entropy $\Delta S_{\text{mix}}$, mixing enthalpy $\Delta H_{\text{mix}}$, atomic size difference $\delta$, $\Omega$, $\eta$, $k_1^{\text{cr}}$, $\frac{E_2}{E_0}$, and $\Delta\chi$ that were defined using the equations provided in Table 1[41]:

**Table 1:** Representative equations used to calculate input features of this study

| Equation | Parameters |
|---|---|
| $\Delta S_{\text{mix}} = -R\sum_{i=1}^{N} c_i \ln(c_i)$ | $R$ is the gas constant which is equal to 8.314 J/(mol.T), and $c_i$ is the concentration of element $i$ in atomic fraction. |
| $\Delta H_{\mathrm{mix}} = \sum_{i=1,\ i\neq j}^{N} 4\Delta H_{ij}^{\mathrm{mix}} c_i c_j$ | $\Delta H_{ij}^{\mathrm{mix}}$ are calculated from available tables that were obtained by Miedema's model [42] |
| $\Omega = \frac{T_{\mathrm{m}}\Delta S_{\mathrm{mix}}}{\lvert\Delta H_{\mathrm{mix}}\rvert}$ | $T_{\mathrm{m}}$ is calculated from $\sum_{i=1}^{N} c_i T_i^{\mathrm{m}}$ wherein $T_i^{\mathrm{m}}$ is the melting temperature of element *i* |
| $\eta = \frac{-T_{\mathrm{ann}}\Delta S_{\mathrm{mix}}}{\lvert\Delta H_{\mathrm{f}}\rvert}$ | $T_{\mathrm{ann}}$ is estimated as $0.8T_{\mathrm{m}}$, and $\Delta H_{\mathrm{f}}$ is the most negative binary mixing enthalpy for forming intermetallic phase ( i.e. $H_{ij}^{\mathrm{IM}}$ ) that are reported in [43] |
| $k_1^{\mathrm{cr}} = \frac{(1 - \frac{0.4T_{\mathrm{m}}\Delta S_{\mathrm{mix}}}{\lvert\Delta H_{\mathrm{mix}}\rvert})}{\frac{\Delta H_{\mathrm{IM}}}{\Delta H_{\mathrm{mix}}}}$ | $\Delta H_{\mathrm{IM}}$ is mixing enthalpy for forming intermetallic phase |
| $\delta = 100 \times \sqrt{\sum_{i=1}^{N} c_i[1 - \frac{r_i}{\sum_{j=1}^{N} c_i r_j}]^2}$ | $r_i$ is the atomic radius of element *i*. |
| $\frac{E_2}{E_0} = \sum_{j\geq i}^{N} \frac{c_i c_j \lvert r_i + r_j - 2\bar{r}\rvert^2}{(2\bar{r})^2}$ | $E_2$ represents the elastic strain energy arising from atomic size mismatch among adjacent atomic pairs in a solid solution, $E_0$ is the total elastic interaction energy assuming no size mismatch [44], and $\bar{r}$ is calculated from $\sum_{i=1}^{N} c_i r_i$ |
| $\Delta\chi = \sqrt{\sum_{i=1}^{N} c_i[\chi_i - \sum_{j=1}^{N} c_j \chi_j]^2}$ | $\chi_i$ is Electronegativity of element *i* |

In addition to these parameters in Table 1, other parameters include: the valence electron concentration $VEC$, bulk modulus $K$, melting temperature $T_{\mathrm{m}}$, atomic number, group ( i.e. the vertical columns numbered 1 through 18 in the periodic table), families (i.e. Alkali metals, Alkaline-earth metals, Rate-earth elements, Transition metals, basic metals, semi-metals, Nonmetals, Halogens, Noble gases), quantum number L, miracle radius, covalent radius, Zunger radius, Mulliken-Badger (MB) electronegativity, Gordy electronegativity, boiling point, density, specific heat, temperature $T$, the phase formation parameters (PFP) [30] for different phases ($PFP_{\mathrm{FCC}}$, $PFP_{\mathrm{BCC}}$, $PFP_{\mathrm{HCP}}$, $PFP_{\mathrm{B_2}}$, $PFP_{\mathrm{Laves}}$, and $PFP_{\mathrm{Sigma}}$), the phase separation parameter (PSP) [30], $\Phi$ which quantifies the change in Gibbs free energy for the formation of a solid solution phase, normalized by the lowest possible Gibbs free energy from binary systems, and is calculated using the publicly available Alloy Search and Predict (ASAP) code [45], the standard deviations of mixing enthalpy $\sigma_{\Delta H_{\mathrm{mix}}}$, bulk modulus $\sigma_K$, melting temperature $\sigma_{T_{\mathrm{m}}}$, and valence electron concentration $\sigma_{VEC}$. It should be noted that the mean value of several parameters, including $K$, $T_{\mathrm{m}}$ , $\chi$, atomic number, group, families, quantum number L, miracle radius, covalent radius, Zunger radius, MB electronegativity, Gordy electronegativity, boiling point, density, and specific heat are calculated as $x_{\mathrm{avg}} = \sum_{i=1}^{N} c_i x_i$ where $c_i$ is the concentration of element $i$ in atomic fraction and $x_i$ is the parameter value for element $i$. The values of $x_i$ are publicly available [32,46]. This reduced set of features captures a wide range of physicochemical properties relevant to phase prediction of RMPEAs while minimizing redundancy and optimizing feature distribution. The complete list of all 51 features is given in the Supplementary Materials. In calculating features using the mean mixture rule, all elements are assumed to be in their stable reference states. Also, the phase formation parameter (PFP) and the phase separation parameter (PSP) are defined as the probability of forming different phases and phase separation respectively [30].

Regarding the architecture of the DL model, we employ an MLP neural network, as illustrated in Figure 1. The hyperbolic tangent (tanh) activation function is used for all layers except the final one, which uses the softmax activation function. The softmax activation function is used to ensure that the sum of the predicted phase fractions is equal to 1, which is a physical constraint in our problem.

To optimize the network's performance, we utilize Bayesian optimization [49,50] for tuning several hyperparameters, including the number of hidden layers, the number of neurons in each layer, batch size, learning rate, the inclusion of batch normalization and dropout layers, as well as the dropout rate. The optimized hyperparameters include a dropout rate of 0.29791, batch size of 1000, 11 hidden layers with 86 neurons in each layer, and learning rate of 0.00065, with batch normalization implemented. These settings balance the model's ability to learn complex patterns while preventing overfitting. More details of our Bayesian optimization process for hyperparameter tuning are provided in the Supplementary Materials.

For training, we utilize the mean square error (MSE) loss function in conjunction with Adam optimizer [51], which adapts the learning rate during training for more efficient convergence. To further avoid overfitting and ensure robust performance, we implement standard 10-fold cross-validation [52,53]. In this approach, 10% of the labeled dataset is first randomly selected as the testing set, while the remaining 90% is divided into ten subsets. For each iteration of the 10-fold cross-validation, nine subsets are used for training the model, and the remaining subset is used for validation. This process is repeated for all subsets, ensuring each subset is considered as the validation set once. The model's accuracy is then averaged over these ten folds.

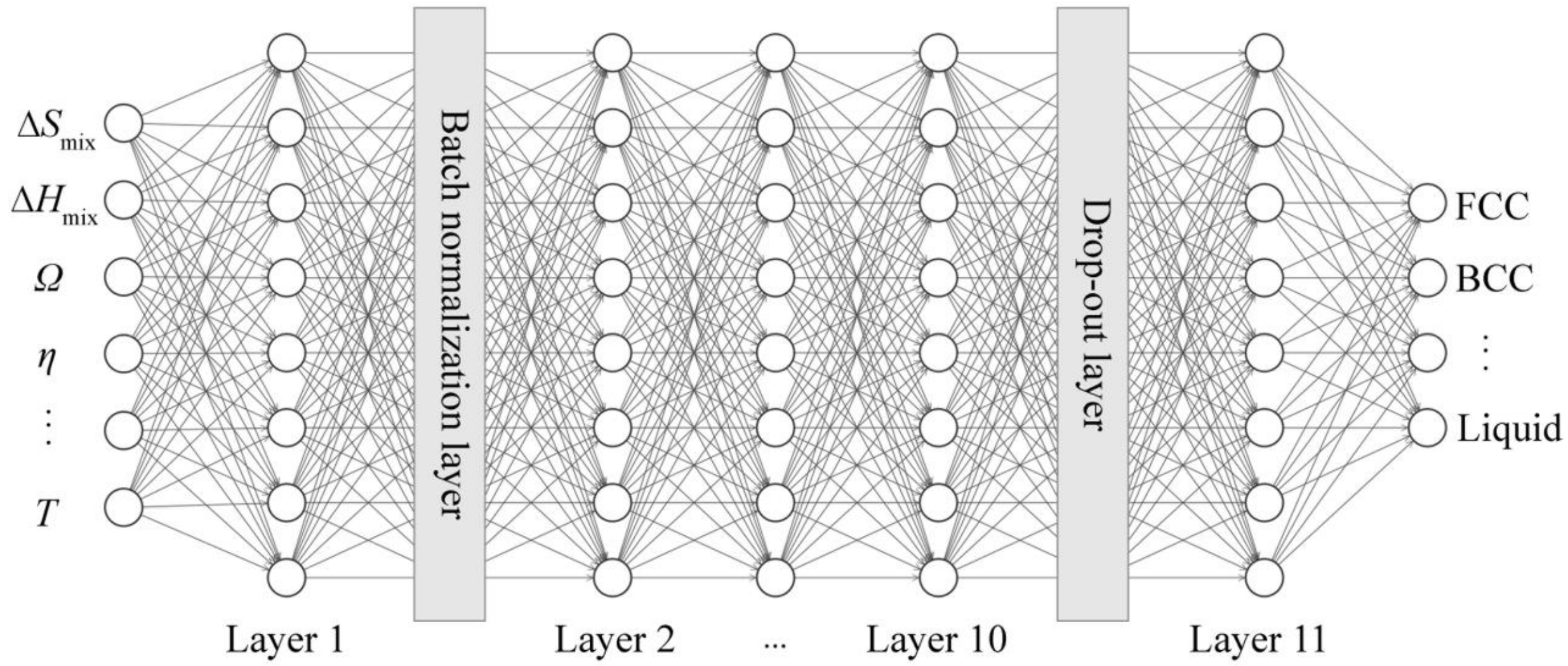

**Figure 1:** The architecture of the MLP network in this study. In this DL model, there are 35 input features, 8 output labels, and 11 hidden layers each having 86 neurons.

# 3. Results

## 3.1. Validation of framework accuracy

The performance of our MLP model during the training process is shown in Figure 2(a), which shows the learning curve with the root mean square error (RMSE) as the performance matric. The model is trained to predict the fractions of eight constituent phases in the RMPEA dataset. The learning curves show that during the first 200 epochs, the RMSE of the training dataset decreases rapidly before plateauing. Concurrently, the RMSE of the validation dataset follows a similar trend, reaching its minimum value at epoch 1406. This convergence indicates that the DL model has successfully captured the complex relationship between the input features and the output RMPEA phases. The best-performing model at Epoch 1406 achieves RMSE of 0.00044 ± 0.0000536, 0.00084 ± 0.0000718, and 0.00089 ± 0.0000586 on the training, validation, and testing datasets, respectively. Here, "±" represents the standard deviation across the 10 cross-fold validation sets. Importantly, to prevent data leakage between the training/validation and unseen separate test datasets, we have carefully evaluated the randomly sampled 50,000 unique compositions from the whole pool of 12 million RMPEAs to ensure that duplicate compositions are not included in our dataset. Moreover, we have also analyzed the Euclidean distances between the compositions in the testing dataset and their nearest neighbors in the training/validation dataset. The histogram of these distances is shown in Supplementary Figure S1. The average Euclidean distance is approximately 0.031, with a minimum of 0.01. This confirms that the compositions of our testing dataset are sufficiently distinct from the training set, reducing the risk of data leakage and ensuring a more robust model evaluation.

Given the complexity of reporting accuracy in multilabel regression problems, we introduce an 'absolute deviation tolerance per phase' metric, which provides a more intuitive measure of our model's accuracy. This metric is defined as the maximum allowable difference between the predicted and true values for each phase in the alloy system. Since both true and predicted values are fractional numbers between 0 and 1, the absolute deviation tolerance per phase is unitless as it is a direct comparison on the same scale as the true values. This metric is shown in Figure 2(b), where each point represents the model's accuracy when the predicted fractions of all 8 phases fell within a specific deviation tolerance compared to true phase fractions. Notably, the figure shows that the DL model can predict the fractions of all 8 phases within a deviation tolerance of 0.07 with an accuracy of 0.9. We also separately report this metric for ternary, quaternary, and quinary RMPEA groups in Figure 2(b), and observe only minor variations between these subgroups and the overall dataset.

Remarkably, such a high level of accuracy is achieved across a wide temperature range. This represents a significant advancement over earlier studies, which were constrained to specific temperatures, as was highlighted in the introduction.

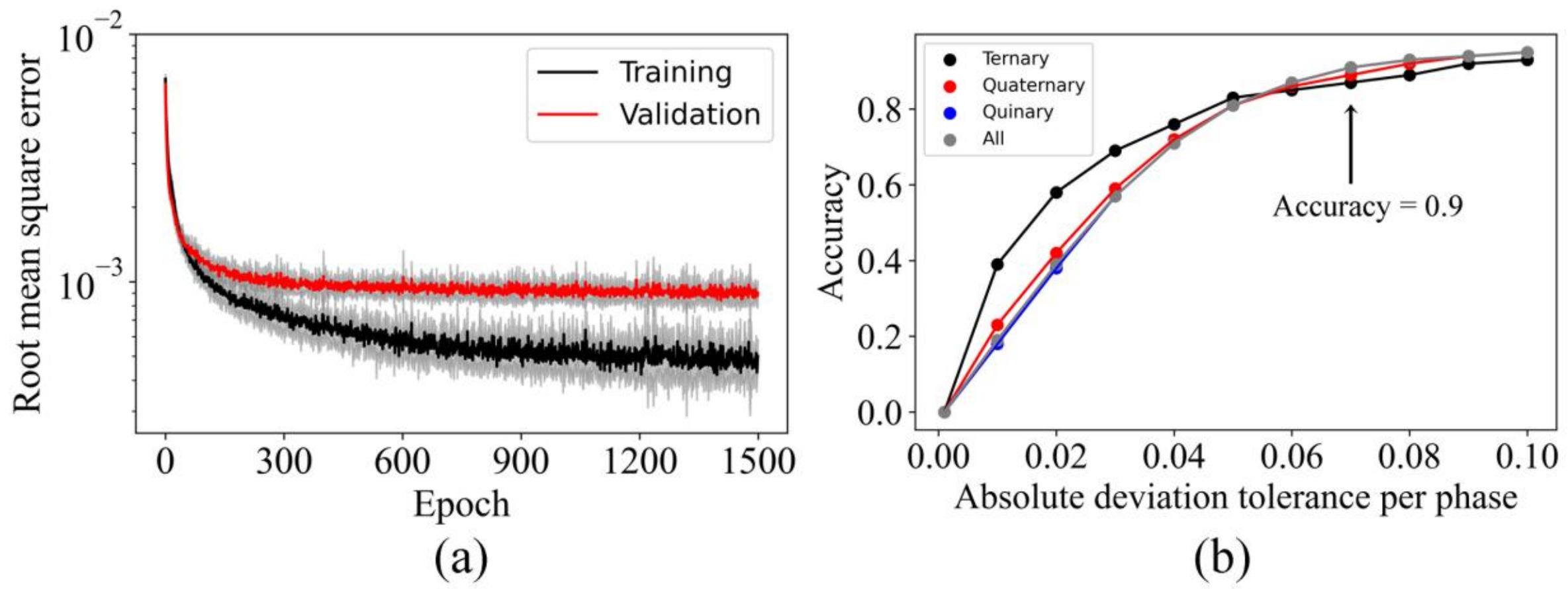

**Figure 2:** (a) Learning curves showing RMSE for training and validation datasets over epochs, with optimal performance at epoch 1406. The vertical axis is plotted on a logarithmic scale. The solid lines represent the average training accuracy (black) and validation accuracy (red) across the ten folds, while the shaded gray areas depict the accuracy variation among individual folds. (b) Relationship between absolute deviation tolerance and prediction accuracy of the DL model (fold=1) for all 8 phases in the testing dataset at epoch 1406 for ternary, quaternary, quinary, and all RMPEAs. Note that our testing dataset has 243 ternary RMPEAs, 4520 quaternary, and 29695 quinary. The x-axis represents the absolute deviation tolerance, while the y-axis shows the corresponding accuracy achieved.

To further quantify the performance of our DL model, the comparison between the true and predicted phase fractions for all 8 phases in the testing dataset are shown as parity plots in Figure 3. These plots show high accuracy across all phases, with most points clustering tightly around the diagonal 45° line. As listed in each subfigure, the coefficient of determination, $R^2$, for each phase further corroborates this observation. The high reported $R^2$ of 0.95 for all phases, except for HCP with an $R^2$ of 0.88, underscores the model's robust performance in predicting RMPEA phase fractions across various temperatures. It should be noted that the relatively lower $R^2$ value for the HCP phase as compared to other phases is due to the low number of data points that have a non-zero fraction of HCP, as observed in Figure 3. In other words, the dataset is more imbalanced with respect to this phase. It is noteworthy that recent studies [31,47,48] have proposed using DFT-calculated binary mixing enthalpies combined with the regular solution model to estimate the mixing enthalpy of BCC RMPEAs. This approach arises from concerns that the Miedema model was originally developed for liquid alloys and can exhibit significant deviations from values observed in crystalline binary systems. Motivated by this, we also evaluated DFT-based mixing enthalpies [47]. Although this led to a modest reduction in model accuracy, the overall trends and predictions remained unchanged. These results suggest that while DFT-based enthalpies may be more physically accurate, Miedema-derived values enable the deep learning model to learn more effective classification boundaries.

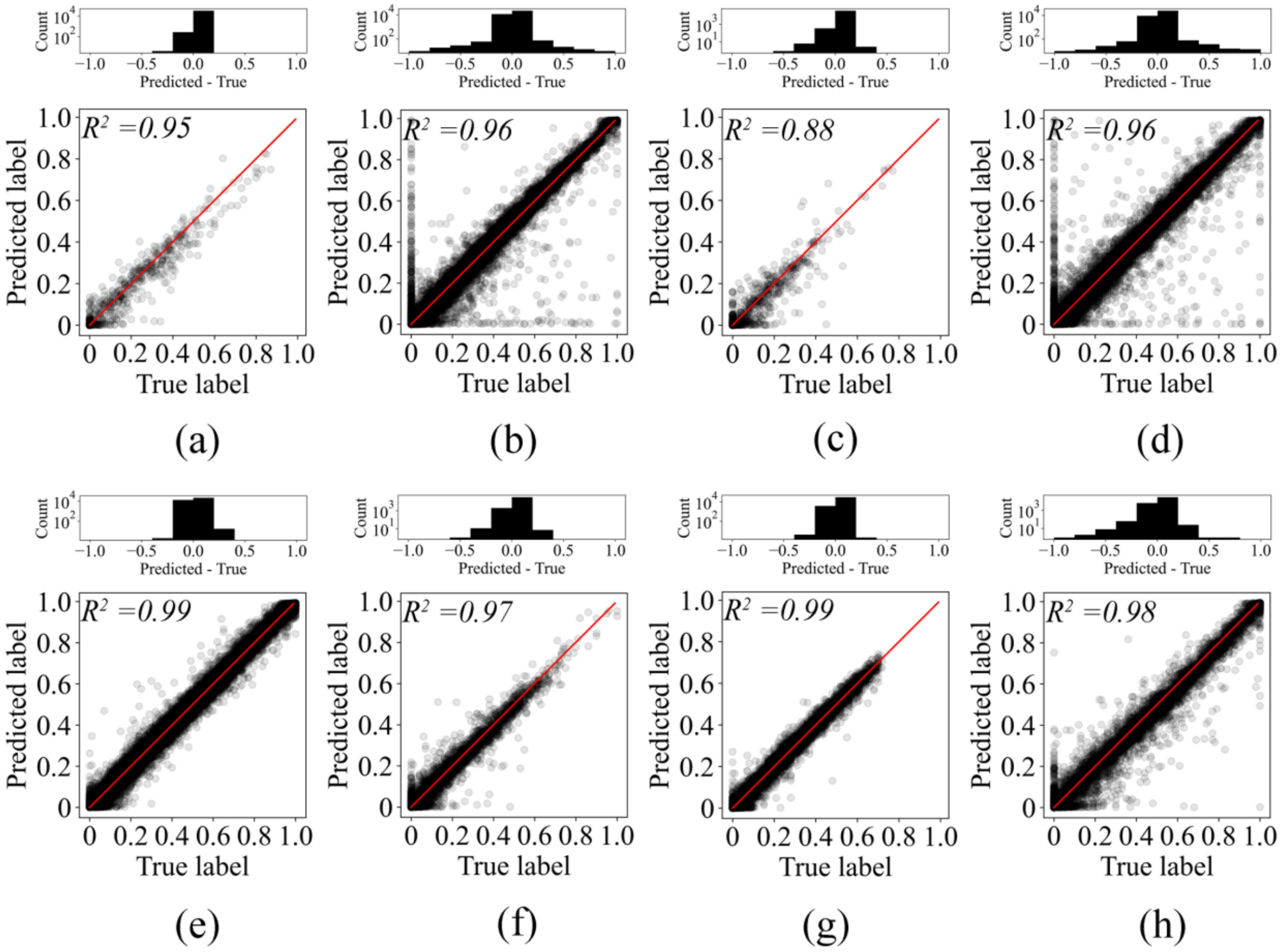

**Figure 3:** Parity plots of the predicted and true phase fraction of the DL model (fold = 1) for the testing dataset. Each subplot represents one of the eight phases: (a) FCC, (b) BCC, (c) HCP, (d) B2, (e) Laves, (f) sigma, (g) Heusler, and (h) liquid. The red diagonal line (y = x) indicates perfect prediction. $R^2$ values for each phase are displayed in their respective subplots. Note that the histogram of the error (i.e., Predicted – True) is shown on top of each subplot, which highlights that most of the data points shown in the parity plots are aligned along the diagonal 45° line.

Additionally, a closer examination of the parity plots reveals an interesting anomaly: the model predicts non-zero values for some RMPEAs that should have true zero BCC or B2 fractions. This is evident in Figure 3(b) and 3(d), where some data points are aligned vertically along the true label = 0 line for these two phases. To investigate this further, we plot the difference between the predicted and true labels (|Pred - True|) for BCC against B2 phases in Figure 4(a). Surprisingly, a subset of RMPEAs closely follow the diagonal 45° line in this plot, indicating a correlation between prediction errors for BCC and B2 phases. Notably, this correlation is unique to the BCC-B2 pair and is not observed in error plots for other phase combinations.

Further analysis shows that for 308 out of 316 RMPEAs having BCC |Pred - True| > 0.1 (lying along the diagonal 45° line in Figure 4(a)) the model accurately predicts the combined BCC + B2 fraction. This suggests that while the model struggles to distinguish between BCC and B2 phases for these alloys and thus occasionally predicts BCC and B2 phases in a flipped manner, it accurately computes their total volume fraction. To test this observation further, we re-train the DL model with BCC and B2 labels merged into one, resulting in a 7-phase output model. The new $R^2$ values are 0.95 for FCC, 0.99 for combined BCC and B2, 0.92 for HCP, 0.99 for Laves, 0.98 for Sigma, 0.99 Heusler, and 0.99 for Liquid phases. Compared with the original DL model, the improved $R^2$ values, particularly for the combined BCC/B2 phase, confirm the model's enhanced performance when treating these phases as a single entity.

It is worth noting that this difficulty in distinguishing between BCC and B2 phases is not unique to our model. Previous studies have reported that Thermo-Calc, the software used to generate our training data, also struggles with distinguishing between these two phases due to their similar Gibbs free energies [54]. Since our DL model was trained on Thermo-Calc outputs, we hypothesize that the observed discrepancies may reflect limitations in the training data. In particular, the model may have learned patterns from the training data's input features and phase labels that inadvertently highlight inconsistencies already present in the original Thermo-Calc data. However, validating this hypothesis would require extensive experimental studies beyond the scope of the current study. Another hypothesis is that there may be additional input features that, if included, could help distinguish the few misclassified samples more effectively.

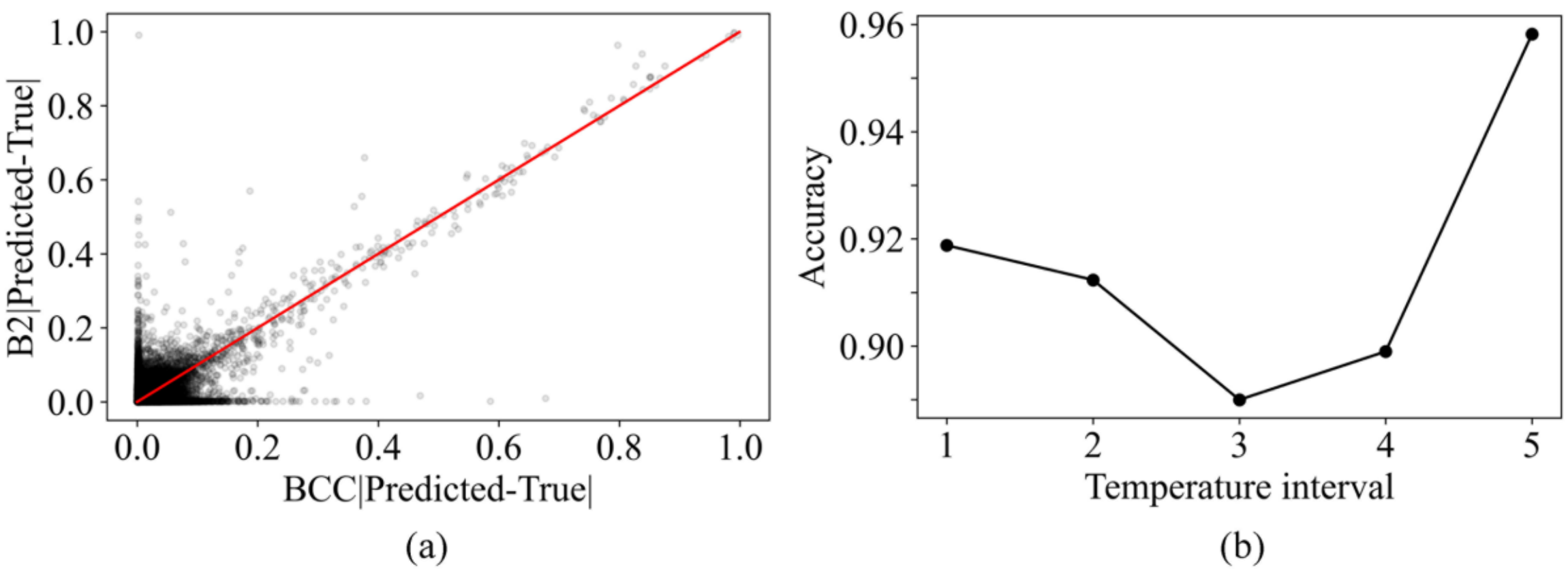

**Figure 4:** (a) Scatter plot showing the correlation between prediction errors of the DL model (fold = 1) for BCC and B2 phases in the testing dataset. The x-axis represents the difference between predicted and true BCC phase fractions, while the y-axis shows the difference between predicted and true B2 phase fractions. The diagonal 45° line indicates equal prediction errors for both phases. (b) Prediction accuracy of the DL model (fold = 1) for all 8 phases using an absolute deviation tolerance of 0.07 across different temperature intervals in the testing dataset at epoch 1406. Note that the temperature intervals are considered as follows: 1) T < 1050 K, 2) 1050 K < T < 1250 K, 3) 1250 K < T < 1450 K, 4) 1450 K < T < 1650 K, and 5) T > 1650 K.

Next, we quantify the performance of our model at different temperatures in a more detailed manner. For this, we divide our testing dataset into several temperature intervals: 1) T < 1050 K, 2) 1050 K < T < 1250 K, 3) 1250 K < T < 1450 K, 4) 1450 K < T < 1650 K, and 5) T > 1650 K. We then calculate the $R^2$ score for each phase within each interval. These results are summarized in Table 2. It is evident from the table that the DL model exhibits strong performance across most temperature intervals, as is indicated by consistently high $R^2$ values. For intervals where the $R^2$ is reported as N/A, this is due to all true labels being zero within those ranges, making the metric undefined. In the case of predicting the HCP phase within the 1250 K < T < 1450 K interval, one should note that the lower $R^2$ value of 0.68 is attributed to the limited variation in the data wherein only 8 out of 8000 samples have non-zero true labels, reducing the reliability of the score in this region. We further evaluate the accuracy of our DL model using an absolute deviation tolerance of 0.07 across different temperature intervals, as shown in Figure 4(b). The results indicate that the model predicts the fractions of all eight phases with an accuracy greater than almost 0.9, across all five temperature intervals.

**Table 2:** $R^2$ performance of the MLP model (fold=1) for phase fraction prediction of RMPEAs across different temperature intervals in the testing dataset

| Phase | T < 1050 K | 1050 K < T < 1250 K | 1250 K < T < 1450 K | 1450 K < T < 1650 K | T > 1650 K |
|---|---|---|---|---|---|
| **FCC** | 0.96 | 0.94 | 0.95 | 0.97 | N/A |
| **BCC** | 0.95 | 0.94 | 0.96 | 0.98 | 0.98 |
| **HCP** | 0.89 | 0.87 | 0.68 | N/A | N/A |
| **B2** | 0.97 | 0.95 | 0.95 | 0.96 | 0.98 |
| **Laves** | 0.99 | 0.99 | 0.99 | 0.98 | 0.97 |
| **Sigma** | 0.97 | 0.96 | 0.95 | 0.95 | N/A |
| **Heusler** | 0.99 | 0.99 | 0.97 | 0.81 | N/A |
| **Liquid** | 0.95 | 0.96 | 0.97 | 0.98 | 0.98 |

## 3.2. Design space exploration for single-phase BCC

Having demonstrated the high accuracy of our DL model in predicting RMPEA phase fractions, in this section, we leverage this model to explore the vast design space for single-phase BCC RMPEAs. Our focus is on predictions at three specific temperatures of 850 K, 1200 K, and 1500 K, which fall within the model's training range. This approach overcomes the limitations of conventional techniques, such as high computational cost, limited compositional coverage, and the need for extensive prior knowledge or trial-and-error screening. By contrast, our model enables rapid, large-scale initial screening of RMPEA design space to identify promising single-phase

BCC candidates across a broad temperature range. This approach also establishes the foundation for developing a multi-fidelity model in future studies. By fine-tuning our current CALPHAD-trained deep learning model using experimental data from collaborators, we aim to integrate the speed of computational predictions with the reliability of experimental measurements.

To achieve this, we discretize the entire design space using elemental composition increments of 3% for each of the three temperatures, resulting in 184,821 unique RMPEA compositions per temperature. We then apply our DL model to predict the phase composition for each of these RMPEAs and identify compositions where BCC is the dominant phase, using various volume fraction thresholds. Here, we consider a range of threshold values, from a strict 0.99 down to a more inclusive 0.8. The reason for exploring a larger threshold of 0.8 is to avoid prematurely excluding promising candidates that may still be of interest to experimentalists, depending on their specific objectives or application requirements. Importantly, while CALPHAD takes 25.99 seconds to label the phases of 15 RMPEA samples, our deep learning model completes the same task in only 0.10 seconds. Consequently, exploring the entire space at each temperature using CALPHAD would require 89 hours, whereas our model accomplishes this in just 20 minutes.

Figure 5(a) shows the number of predicted single-phase BCC RMPEAs at different temperatures for BCC volume fraction thresholds. From this figure, it is observed that the number of single-phase BCC RMPEAs increases with temperature. In addition, as the BCC volume fraction threshold increases, the number of RMPEAs at or above decreases. One should note that our CALPHAD calculations on candidate RMPEAs, based on the BCC volume fraction threshold of 0.90, yields an overall prediction accuracy of 0.95 with respective accuracies of 0.78, 0.96, and 0.95 at 850 K, 1200 K, and 1500 K thereby validating our design space exploration. The reduced accuracy at 850 K may be attributed to the limited number of RMPEAs exhibiting a BCC phase fraction ≥ 0.90 at lower temperatures in the training dataset.

To further characterize the single-phase BCC region of the design space, we visualize the elemental compositions of RMPEAs meeting a BCC volume fraction threshold of 0.90 using a spider plot, as shown in Figure 5(b-d). This figure reveals several interesting trends. First, Ni content is consistently low across all temperatures, suggesting Ni is not desirable in large volume fractions when stable single-phase BCC alloys are desired. Second, Fe and Al show temperature-dependent behavior. At lower temperatures, their content is limited to single-phase BCC alloys. On the other hand, their content can increase substantially at higher temperatures. This suggests that Fe and Al act as high-temperature stabilizers for single-phase BCC structure. The capability of Al to stabilize single-phase BCC has been demonstrated in previous studies as well [55,56]. These insights provided valuable guidance for designing new single-phase BCC RMPEAs in future experimental studies. By understanding the role of each element and how it changes with temperature, researchers can more efficiently target compositions likely to yield the desired single-phase BCC structure.

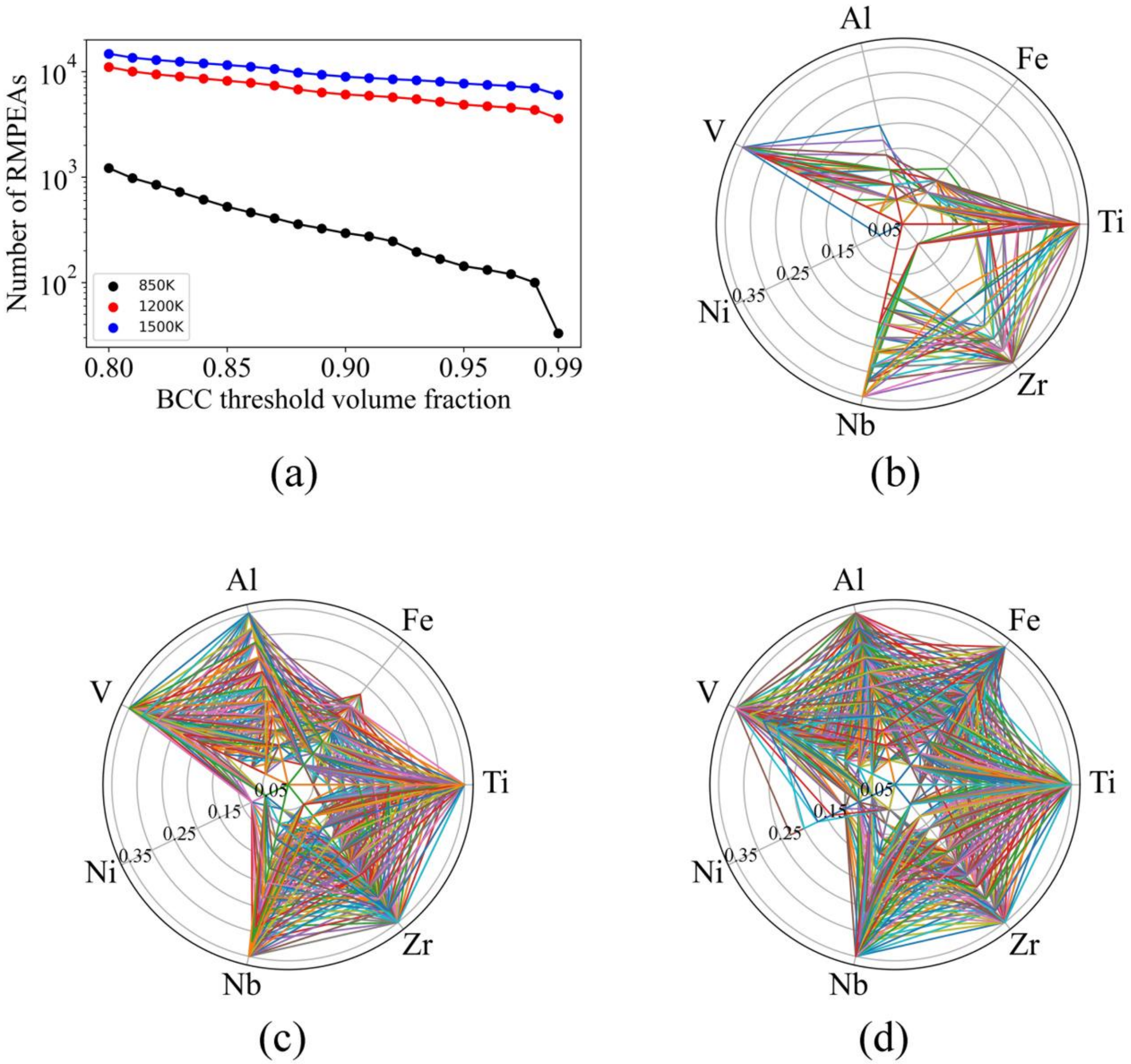

**Figure 5:** Analysis of single-phase BCC RMPEAs predicted using our DL model (fold=1) during the design space exploration. (a) The number of predicted single-phase BCC RMPEAs at 850 K, 1200 K, and 1500 K versus the BCC volume fraction thresholds as predicted during the design space exploration. The spider plot of the elemental distributions of the predicted RMPEAs with BCC volume fraction ≥ 0.9 are shown in (b) at 850K, (c) at 1200K, and (d) at 1500K. Each radial direction represents an element, with distance from the center indicating atomic percentage.

## 3.4. A Predictive Mathematical Model for single-phase BCC RMPEAs

While our DL model accurately predicts RMPEA phases using 35 input features with a considerably higher speed than CALPHAD calculations, its complexity still hinders direct interpretation and application. To address this, we further develop a mathematical model for identifying single-phase BCC RMPEAs, balancing accuracy with interpretability.

To achieve this, we re-label the initial dataset as 0 for non-single-phase BCC and 1 for single-phase BCC. In this re-labeling process, only compositions with 100% BCC phase fraction are considered single-phase BCC. We then conduct a feature importance analysis by training a random forest classifier using scikit-learn [57] to identify key distinguishing features. The random forest classifier is trained on a dataset of 35 input features and categorical class labels using n_estimators = 100 and random_state = 42 with all other hyperparameters set to their scikit-learn defaults. The data is split into training (90%) and validation (10%) subsets using train_test_split. Upon training, the model achieves an overall accuracy of 99% on the validation dataset, with class-wise F1 scores [33] of 0.99 for class 0 and 0.91 for class 1. Importantly, the initial testing dataset's class-wise F1 scores are 0.99 for class 0 and 0.92 for class 1. This confirms the trained random forest model's ability to distinguish key differences in the data. We emphasize that this initial unseen testing dataset is the same as the one used in earlier sections for evaluating the performance of our MLP model. Feature importances extracted from the trained model are then used to rank the most important features. The ranking of these features based on importance is shown in Figure 6(a). It is important to note that the choice of the random forest classifier is due to its ability to inherently provide insights into feature importance straightforwardly and robustly, especially in high-dimensional spaces, compared to the complex interpretability of the MLP model.

We then train a Support Vector Classification (SVC) model to derive a mathematical equation using the top 21 key parameters as input features. We select 21 features because using fewer features leads to a lower F1 score, while including more (up to all 35) offers no significant improvement. Importantly, we choose SVC because it provides a straightforward framework for directly obtaining the coefficients needed to calibrate our mathematical equation. In contrast, the complexity of MLP makes it challenging to extract explicit coefficients for mathematical representation. The SVC model is implemented using a linear kernel and class_weight = 'balanced' to account for class imbalance with a regularization parameter of C=100. At the same time, all other hyperparameters are set to their scikit-learn defaults. The use of class_weight = 'balanced' helps mitigate potential bias in predictions due to the uneven distribution of sample types. The training and validation datasets, initially partitioned for training the random forest, are subsequently used to train the SVC. The SVC was initially trained on the entire training dataset. However, its performance was suboptimal, with class-wise F1 scores of 0.95 for class 0 and 0.43 for class 1 on the initial testing set. Interestingly, the true positives, false positives, false negatives, and true negatives for the initial testing dataset were 676, 1732, 12, and 32038, respectively, indicating that the low F1 score was primarily due to the large number of false positives. Such a low F1 score is not surprising, given the linear nature of the SVC model and the imbalance in our training dataset, where only 6,002 out of 279,112 samples belong to class 1. Additionally, the inherent complexity of the data had previously required a deep random forest to achieve successful classification in the first part of this section.

To improve the model prediction ability, we have retrained the SVC model using all data points belonging to class 1 from the training dataset and the false positive samples also taken from the training dataset. This approach resembles the hard negative mining method [58], where the model is retrained using misclassified negative samples to enhance its discriminative capability. After this training, the SVC model achieves an overall 98% accuracy on the initial testing dataset with a class-wise F1 score of 0.99 for class 0 and 0.72 for class 1. While not perfect, the performance on class 1 is sufficient to capture the boundary between classes 1 and 0, making the model suitable for deriving an interpretable mathematical equation. From this predictive SVC model, we extract the following relationship:

$$F = 60.90 + 5.02 \times PFP_{\mathrm{BCC}} - 31.96 \times \delta - 3.66 \times T - 60.86 \times PFP_{\mathrm{FCC}} - 11.25 \times \eta + 15.03 \times \sigma_{VEC} - 15.91 \times \sigma_{\Delta H_{\mathrm{mix}}} - 5.36 \times PSP + 7.60 \times \Phi - 19.64 \times \Delta H_{\mathrm{mix}} - 36.26 \times \Delta\chi + 23.32 \times PFP_{\mathrm{Laves}} - 16.43 \times R_{\mathrm{Mir}} + 1.64 \times k_1 - 28.65 \times PFP_{\mathrm{B2}} - 1.97 \times PFP_{\mathrm{Sigma}} - 16.77 \times PFP_{\mathrm{HCP}} - 1.38 \times VEC + 4.65 \times \Omega - 1.16 \times Grp + 3.50 \times \Delta S_{\mathrm{mix}} \quad (1)$$

Wherein $R_{\mathrm{Mir}}$ represents the Miracle Radius, and Grp denotes the group number. In this relationship, $F > 0$ indicates a single-phase BCC RMPEA. It should be noted that all input parameters in this equation are normalized to the range [0, 1], with the corresponding minimum and maximum values listed in Supplementary Table S3.

To visualize the performance of our derived equation, we calculate the $F$ values from Eq. (1) for all samples in the initial testing dataset. Figure 6(b) shows the distribution of $F$ values for all RMPEAs in the testing set. The figure clearly demonstrates that single-phase BCC samples have $F > 0$, validating the effectiveness of our derived equation.

This mathematical model offers a computationally efficient method for identifying potential single-phase BCC RMPEAs without requiring complex DL models or CALPHAD calculations. It provides a valuable tool for rapid initial screening in RMPEA design, potentially accelerating the discovery of new alloys with desired properties. By bridging the gap between complex, black-box

DL models and readily interpretable mathematical formulations, this approach contributes to making RMPEA design more accessible and efficient for researchers and engineers in the field.

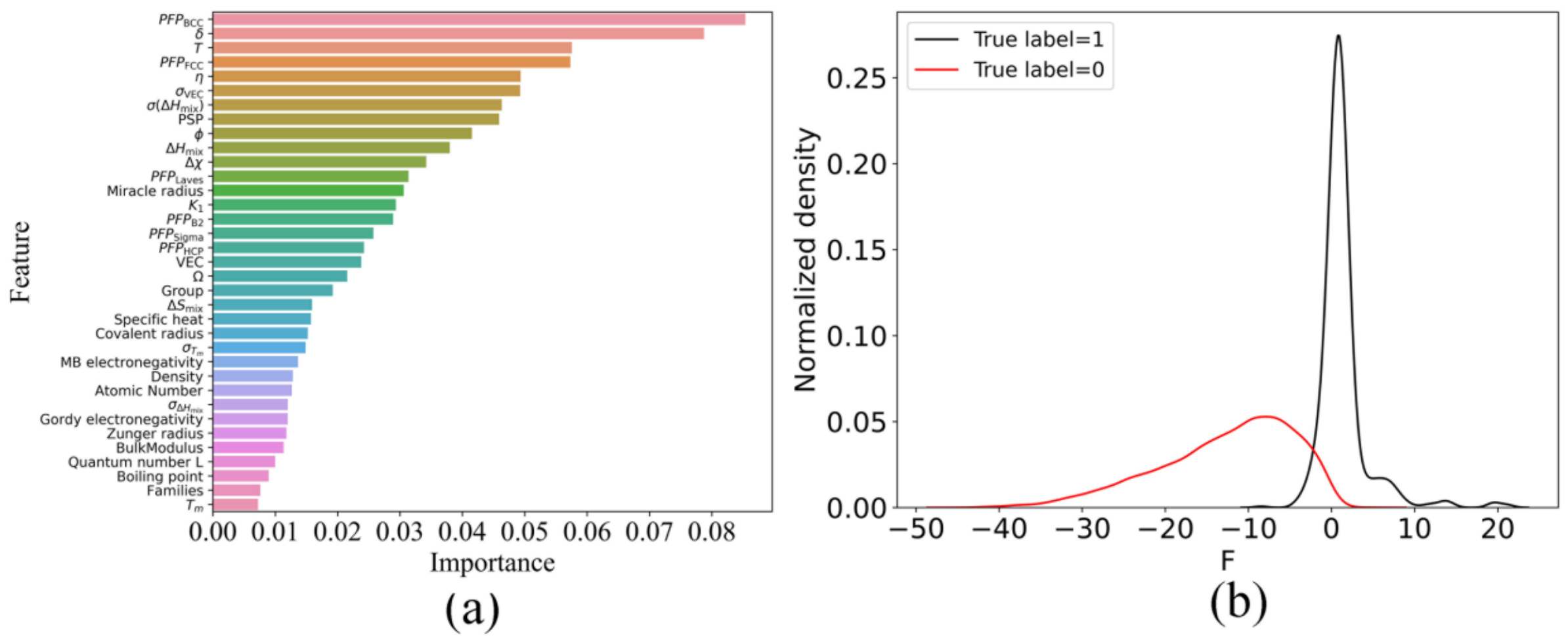

**Figure 6:** (a) Ranking of input features based on random forest importance scores (b)Distribution of the output of the mathematical equation developed in this study for the initial testing dataset.

# 4. Conclusion

In conclusion, we developed a deep learning framework to tackle the ongoing challenge of predicting RMPEA phases across different temperatures. Using the Bayesian optimization technique to refine hyperparameters, an MLP model was designed and trained on a comprehensive dataset of 50,000 RMPEAs. Each alloy was labeled at seven equidistant temperatures from CALPHAD, ranging from 850 K to 0.8 $T_m$ resulting in a total of 350,000 labeled samples within the elemental space of Ti, Fe, Al, V, Ni, Nb, and Zr. The model's performance was thoroughly evaluated using a ten-fold cross-validation strategy. Our results revealed that the model can predict the fraction of up to eight constituent phases of RMPEAs with an accuracy of 0.9 and a deviation tolerance of 0.07 per phase across different temperatures emphasizing the model's robustness and ability for generalization. We further demonstrated that while the model achieved high accuracy in predicting the combined BCC + B2 phase fraction, the accuracy decreased when predicting each phase separately for a group of RMPEAs and thus the DL model had lower accuracy in distinguishing between the BCC and B2 phases. Though this discrepancy could indicate potential limitations of the model, it may also suggest necessary revisions in the true labels obtained from Thermo-Calc given that Thermo-Calc is known to face challenges in distinguishing between BCC and B2 phases. We then navigated the vast compositional space of our elements of interest to identify potentially stable single-phase BCC RMPEAs at various temperatures using our DL model. Among the 184,821 RMPEAs evaluated at each temperature, we predicted that 0.01%, 2.16% and 3.78% of them can be stabilized in a single-phase BCC with at least 90% BCC content at 850K, 1200K and 1500K respectively. As a result, the predicted RMPEAs showed a stronger tendency to form single-phase BCC at higher temperatures. Moreover, our model suggested that inclusion of Ni content was correlated with the lack of single-phase BCC at all temperatures while Fe and Al content was correlated with single-phase BCC at higher temperatures. To accelerate the discovery of single-phase BCC RMPEAs without relying on deep learning or CALPHAD calculations, we then proposed a straightforward mathematical equation. This equation, informed by feature importance analysis and insights from the SVC model, enabled the rapid identification of single-phase BCC RMPEAs across different temperatures with sufficient accuracy. The data-driven approach developed in this study is applicable across a range of technologically relevant temperatures and can be readily applied for phase prediction in diverse material systems, thereby accelerating the discovery of novel materials.

# Acknowledgment

CAIMEE research was sponsored by the Army Research Laboratory and was accomplished under Cooperative Agreement Number W911NF-22-2-0014. The authors gratefully acknowledge internal financial support from the Johns Hopkins University Applied Physics Laboratory's Independent Research & Development (IR&D) Program. Computational resources were provided by the Advanced Research Computing at Hopkins (ARCH).

# Data availability

The data and code supporting the findings of this study are available at https://github.com/ashargh1/RMPEAs.

# Declaration of Competing Interest

The authors declare that they have no known competing financial interests or personal relationships that could have appeared to influence the study reported in this paper.

# Author contributions

**Ali K. Shargh**: Writing – original draft, Writing – review and editing, Visualization, Validation, Software, Methodology, Investigation, Formal analysis, Data curation, Conceptualization.

**Christopher D. Stiles**: Writing – review and editing, Methodology, Funding acquisition, Formal analysis, Conceptualization.

**Jaafar A. El-Awady**: Writing – review and editing, Supervision, Resources, Project administration, Methodology, Investigation, Funding acquisition, Formal analysis, Conceptualization.

# Supplementary Material for

# Temperature-Dependent Discovery of BCC Refractory Multi-principal Element Alloys: Integrating Deep Learning and CALPHAD Calculations

Ali K. Shargh[1*], Christopher D. Stiles[1,2], Jaafar A. El-Awady[1*]

[1] Department of Mechanical Engineering, Johns Hopkins University, Baltimore, Maryland 21218, United States

[2] Research and Exploratory Development Department, Johns Hopkins University Applied Physics Laboratory, Laurel, Maryland 20723, United States

## Composition analysis

The histogram of the Euclidean distances between the compositions in the test set and their nearest neighbors in the training set is shown in Supplementary Figure S1. The average Euclidean distance is approximately 0.031, with a minimum of 0.01. This confirms that the test compositions are sufficiently distinct from the training set, reducing the risk of data leakage and ensuring a more robust model evaluation.

[*] Corresponding authors:
Email addresses: ashargh1@jhu.edu (A. K. Shargh), jelawady@jhu.edu (J. A. El-Awady)

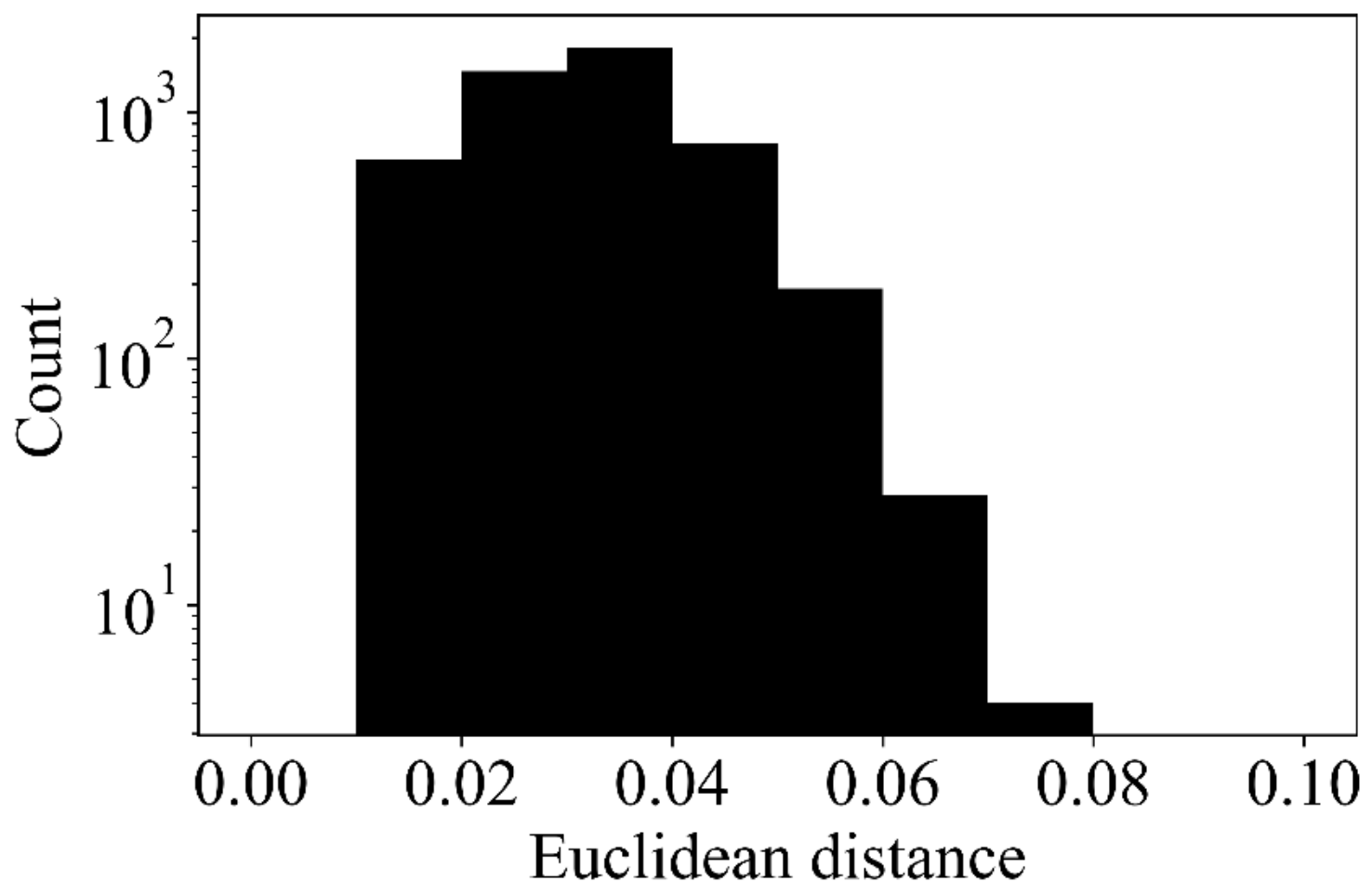

**Figure S1:** Histogram of Euclidean distances between RMPEA compositions of the testing dataset and their nearest neighbors in the training dataset

## Features definition

The list of the 35 reduced input features is provided in the Methods section of the paper. In addition, the complete list of the 51 features initially used in this study, including the 35 retained features, is provided below: mixing entropy $\Delta S_{\mathrm{mix}}$, mixing enthalpy $\Delta H_{\mathrm{mix}}$, $\Omega$, $\eta$, $k_1^{\mathrm{cr}}$, $\Phi$, $\delta$, $\frac{E_2}{E_0}$, valence electron concentration $VEC$, $\Delta\chi$, $PFP_{\mathrm{FCC}}$, $PFP_{\mathrm{BCC}}$, $PFP_{\mathrm{HCP}}$, $PFP_{\mathrm{B_2}}$, $PFP_{\mathrm{Laves}}$, $PFP_{\mathrm{Sigma}}$, $PSP$, $\sigma_{\Delta H_{\mathrm{mix}}}$, bulk modulus $K$, $\sigma_K$, melting temperature $T_{\mathrm{m}}$, $\sigma_{T_{\mathrm{m}}}$, $\sigma_{VEC}$, $\chi$, atomic number, atomic weight, period, group, families, Mendeleev number, L quantum number, miracle radius, covalent radius, Zunger radius, ionic radius, crystal radius, MB electronegativity, Gordy electronegativity, Mulliken electronegativity, Allred-Rockow electronegativity, first ionization potential, polarizability, boiling point, density, specific heat, heat of fusion, heat of vaporization, thermal conductivity, heat atomization and cohesive energy. The phase formation parameter (PFP) and phase separation parameter (PSP) are defined as the probability of forming different phases as well as phase separation respectively [1]. The parameters $\Delta S_{\text{mix}}$, $\Delta H_{\text{mix}}$, $\Omega$, $\eta$, $k_1^{\text{cr}}$, $\delta$, $\frac{E_2}{E_0}$, $\Delta\chi$ are defined in the Supplementary Table S1. The mean value of the remaining parameters from bulk modulus to cohesive energy are calculated from:

$$x_{\text{avg}} = \sum_{i=1}^{N} c_i x_i \tag{1}$$

Wherein $c_i$ is the concentration of element $i$ in atomic fraction, and $x_i$ is the parameter values for element $i$. The values of these parameters are publicly available [2].

**Table S1:** Definition of the input features used in this study. The parameters are shown in the left column while their corresponding descriptions are shown in the right column.

| Parameter | Description |
|---|---|
| $\Delta S_{\text{mix}} = -R\sum_{i=1}^{N} c_i \ln(c_i)$ | $R$ is the gas constant which is equal to 8.314 J/(mol.T), and $c_i$ is the concentration of element $i$ in atomic fraction. |
| $\Delta H_{\text{mix}} = \sum_{i=1,\ i\neq j}^{N} 4\Delta H_{ij}^{\text{mix}} c_i c_j$ | $\Delta H_{ij}^{\text{mix}}$ are calculated from available tables that are obtained from Miedema's model. |
| $\Omega = \frac{T_{\text{m}}\Delta S_{\text{mix}}}{\lvert\Delta H_{\text{mix}}\rvert}$ | $T_{\text{m}}$ is calculated from $\sum_{i=1}^{N} c_i T_i^{\text{m}}$ wherein $T_i^{\text{m}}$ is the melting temperature of element $i$. |
| $\eta = \frac{-T_{\text{ann}}\Delta S_{\text{mix}}}{\lvert\Delta H_{\text{f}}\rvert}$ | $T_{\text{ann}}$ is estimated as $0.8T_{\text{m}}$, and $\Delta H_{\text{f}}$ is the most negative binary mixing enthalpy for forming IM ( i.e. $H_{ij}^{\text{IM}}$ ) that are reported in [3]. |
| $k_1^{\text{cr}} = \frac{(1 - \frac{0.4T_{\text{m}}\Delta S_{\text{mix}}}{\lvert\Delta H_{\text{mix}}\rvert})}{\frac{\Delta H_{\text{IM}}}{\Delta H_{\text{mix}}}}$ | $\Delta H_{\text{IM}}$ is mixing enthalpy for forming intermetallic phase. |
| $\delta = 100 \times \sqrt{\sum_{i=1}^{N} c_i[1 - \frac{r_i}{\sum_{j=1}^{N} c_j r_j}]^2}$ | $r_i$ is the atomic radius of element $i$. |
| $\frac{E_2}{E_0} = \sum_{j\geq i}^{N} \frac{c_i c_j \lvert r_i + r_j - 2\bar{r}\rvert^2}{(2\bar{r})^2}$ | $\bar{r}$ is calculated from $\sum_{i=1}^{N} c_i r_i$. |
| $\Delta\chi = \sqrt{\sum_{i=1}^{N} c_i[\chi_i - \sum_{j=1}^{N} c_j \chi_j]^2}$ | $\chi_i$ is Electronegativity of element $i$. |

We provide the Pearson correlation coefficient matrix of the 51 input features used during the feature engineering step in Supplementary Figure S2, to offer the reader a clearer understanding of the relationships among the features utilized in our model.

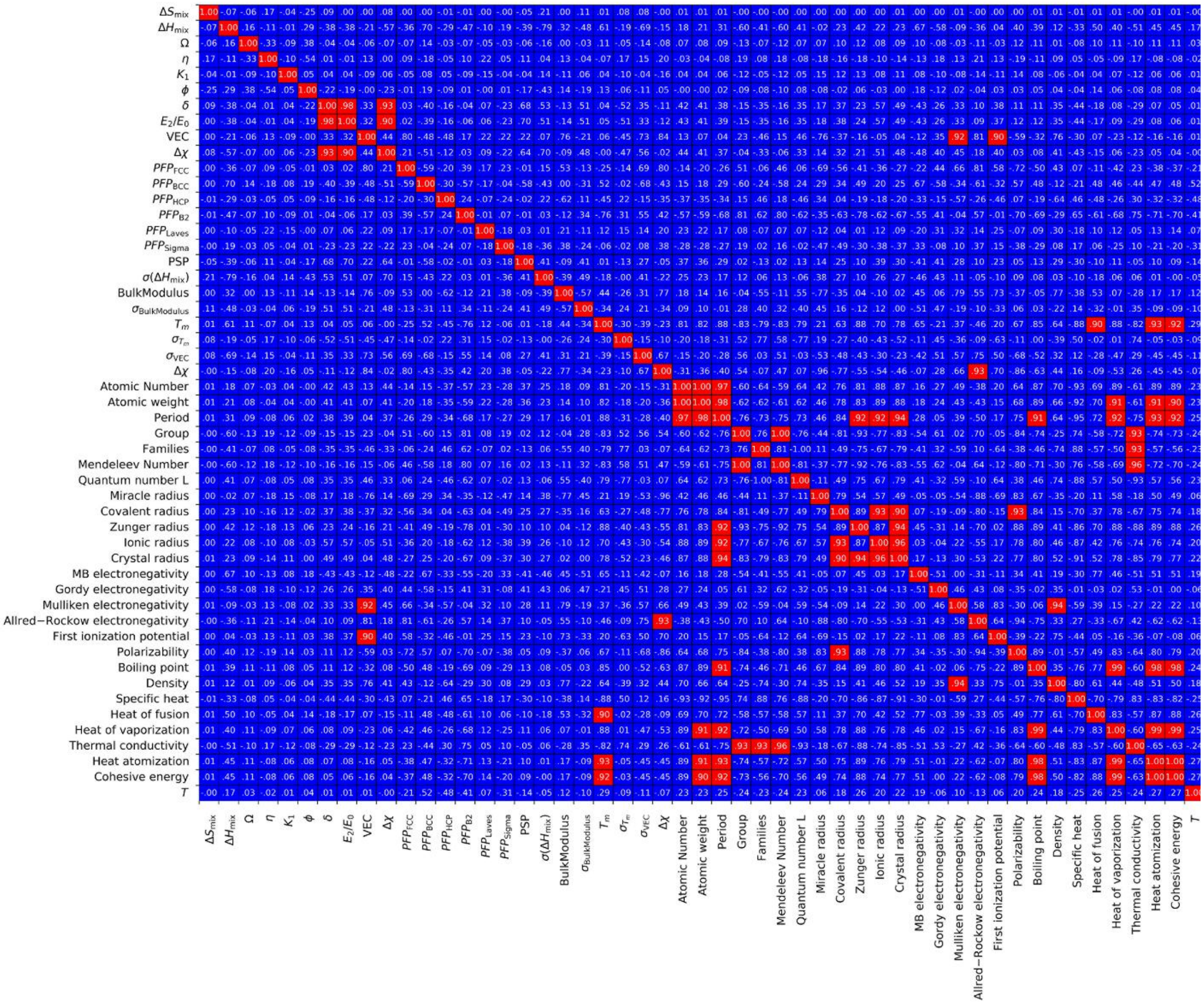

**Figure S2:** Pearson correlation coefficient matrix for the 51 input features (IDs 0 to 50). Note that those pairs that are highly correlated (PCC>0.9) are colored with red. A high-resolution version of this figure is available for clarity and readers are encouraged to zoom in for clarity due to the large number of annotated values in the matrix.

## Bayesian optimization for neural network tuning

To optimize the neural network architecture for predicting mechanical properties of RMPEAs, we employ Bayesian Optimization. The search space includes batch size (32–1024), number of neurons per layer (20–100), number of layers before the drop-out layer (5–10), number of layers after the drop-out layer (0-10), batch normalization layer (on/off), drop-out layer (on/off), drop-out rate (0.1–0.6), and learning rate (0.0001-0.01). The objective function is to minimize the mean square error (MSE), evaluated on a held-out test set. We run 40 random initial evaluations followed by 80 iterations of optimization. This process improves the model's performance. Comparative results for various configurations are summarized in Table S2.

**Table S2:** The results of our Bayesian optimization. Note that a value of 0 or 1 for Drop-out (DO) and batch normalization (BN) layers indicates their absence or presence respectively

| DO | DO rate | BN | Batch size | L1 | L2 | Neurons | Learning rate | MSE |
|---|---|---|---|---|---|---|---|---|
| 0 | 0.31803 | 0 | 639 | 9 | 3 | 22 | 0.00158 | 0.00241 |
| 0 | 0.59536 | 0 | 269 | 6 | 1 | 70 | 0.00673 | 0.00381 |
| 0 | 0.13698 | 1 | 494 | 10 | 8 | 85 | 0.00842 | 0.0413 |
| 1 | 0.31066 | 1 | 605 | 5 | 5 | 85 | 0.00114 | 0.00148 |
| 0 | 0.59924 | 1 | 593 | 6 | 6 | 52 | 0.0049 | 0.00206 |
| 0 | 0.25948 | 1 | 352 | 10 | 9 | 23 | 0.00816 | 0.03772 |
| 1 | 0.34064 | 0 | 975 | 10 | 4 | 21 | 0.00326 | 0.00369 |
| 0 | 0.11669 | 0 | 85 | 6 | 5 | 34 | 0.00187 | 0.00258 |
| 0 | 0.48412 | 1 | 710 | 8 | 4 | 57 | 0.00599 | 0.00228 |
| 1 | 0.34896 | 0 | 179 | 8 | 7 | 89 | 0.00186 | 0.00368 |
| 0 | 0.28075 | 1 | 101 | 7 | 2 | 96 | 0.00184 | 0.00258 |
| 1 | 0.5209 | 0 | 166 | 5 | 2 | 74 | 0.00874 | 0.01335 |
| 1 | 0.20525 | 1 | 761 | 9 | 1 | 37 | 0.00315 | 0.00191 |
| 1 | 0.15759 | 0 | 934 | 10 | 1 | 92 | 0.00633 | 0.04303 |
| 1 | 0.28699 | 0 | 680 | 9 | 9 | 32 | 0.00624 | 0.00392 |
| 0 | 0.36612 | 0 | 663 | 7 | 7 | 56 | 0.00482 | 0.0029 |
| 0 | 0.27337 | 1 | 441 | 10 | 2 | 23 | 0.00686 | 0.0035 |
| 1 | 0.39591 | 0 | 728 | 6 | 7 | 53 | 0.00041 | 0.00292 |
| 0 | 0.52029 | 1 | 137 | 6 | 10 | 78 | 0.00241 | 0.00621 |
| 1 | 0.23265 | 0 | 109 | 10 | 8 | 86 | 0.00405 | 0.04287 |
| 0 | 0.56292 | 1 | 550 | 9 | 5 | 21 | 0.00126 | 0.00264 |
| 1 | 0.14742 | 0 | 540 | 9 | 1 | 98 | 0.00639 | 0.04333 |
| 0 | 0.54493 | 0 | 686 | 10 | 7 | 40 | 0.00645 | 0.00285 |
| 1 | 0.23344 | 0 | 298 | 10 | 2 | 79 | 0.00305 | 0.00329 |
| 1 | 0.34913 | 1 | 62 | 6 | 0 | 63 | 0.00852 | 0.008 |
| 1 | 0.2941 | 0 | 688 | 7 | 8 | 86 | 0.00474 | 0.04278 |
| 0 | 0.15904 | 0 | 426 | 6 | 7 | 50 | 0.00931 | 0.04331 |
| 1 | 0.11249 | 1 | 898 | 9 | 6 | 67 | 0.00521 | 0.00251 |
| 0 | 0.22909 | 0 | 204 | 5 | 6 | 93 | 0.00358 | 0.00889 |
| 0 | 0.23676 | 0 | 136 | 9 | 4 | 55 | 0.00801 | 0.04299 |
| 0 | 0.43142 | 1 | 43 | 8 | 6 | 79 | 0.00498 | 0.04284 |
| 0 | 0.57007 | 1 | 807 | 9 | 7 | 74 | 0.00938 | 0.04225 |
| 0 | 0.38967 | 0 | 374 | 5 | 6 | 74 | 0.00636 | 0.00881 |
| 0 | 0.22574 | 1 | 604 | 10 | 9 | 90 | 0.00132 | 0.00155 |
| 0 | 0.38857 | 1 | 488 | 7 | 2 | 40 | 0.00652 | 0.00239 |
| 1 | 0.46465 | 0 | 925 | 6 | 4 | 79 | 0.00262 | 0.00202 |
| 1 | 0.50653 | 1 | 478 | 5 | 3 | 51 | 0.00583 | 0.0039 |
| 0 | 0.48221 | 1 | 557 | 6 | 4 | 96 | 0.0001 | 0.00177 |
| 1 | 0.25742 | 0 | 125 | 6 | 3 | 89 | 0.00738 | 0.04342 |
| 1 | 0.30266 | 0 | 44 | 9 | 1 | 22 | 0.00843 | 0.04343 |
| 1 | 0.19858 | 1 | 808 | 8 | 7 | 73 | 0.00419 | 0.0024 |
| 0 | 0.15155 | 1 | 906 | 10 | 1 | 22 | 0.00516 | 0.00251 |
| 1 | 0.47598 | 1 | 242 | 6 | 8 | 81 | 0.00254 | 0.00369 |
| 1 | 0.25876 | 0 | 695 | 7 | 6 | 85 | 0.00493 | 0.00784 |
| 1 | 0.45664 | 0 | 682 | 6 | 9 | 50 | 0.00884 | 0.04279 |
| 1 | 0.5352 | 1 | 953 | 8 | 3 | 86 | 0.0036 | 0.0021 |
| 0 | 0.31244 | 1 | 77 | 10 | 9 | 47 | 0.00562 | 0.04391 |
| 0 | 0.24537 | 0 | 101 | 8 | 0 | 47 | 0.00107 | 0.00206 |
| 0 | 0.40201 | 0 | 744 | 6 | 6 | 42 | 0.00739 | 0.00252 |
| 0 | 0.34052 | 1 | 270 | 5 | 0 | 85 | 0.00109 | 0.00141 |
| 0 | 0.30206 | 1 | 268 | 6 | 9 | 35 | 0.00246 | 0.00228 |
| 1 | 0.24245 | 0 | 875 | 8 | 3 | 49 | 0.00167 | 0.00192 |
| 1 | 0.24683 | 1 | 210 | 6 | 5 | 35 | 0.00384 | 0.00318 |
| 0 | 0.40924 | 1 | 804 | 8 | 7 | 60 | 0.00024 | 0.00178 |
| 0 | 0.35027 | 0 | 533 | 8 | 8 | 72 | 0.00383 | 0.04275 |
| 1 | 0.24143 | 1 | 908 | 10 | 3 | 62 | 0.00943 | 0.01706 |
| 0 | 0.17638 | 0 | 563 | 8 | 9 | 48 | 0.00658 | 0.0428 |
| 0 | 0.2143 | 1 | 913 | 9 | 5 | 76 | 0.00266 | 0.00178 |
| 1 | 0.5846 | 0 | 477 | 7 | 10 | 37 | 0.00954 | 0.04292 |
| 0 | 0.12791 | 0 | 217 | 9 | 9 | 49 | 0.00218 | 0.00245 |
| 0 | 0.52794 | 0 | 759 | 6 | 5 | 43 | 0.00982 | 0.00259 |
| 1 | 0.53903 | 0 | 679 | 10 | 0 | 44 | 0.00043 | 0.00267 |
| 0 | 0.34642 | 0 | 976 | 6 | 2 | 73 | 0.00957 | 0.00264 |
| 0 | 0.44406 | 0 | 237 | 9 | 4 | 41 | 0.00288 | 0.00254 |
| 0 | 0.44976 | 1 | 699 | 7 | 2 | 49 | 0.00441 | 0.00243 |
| 0 | 0.25245 | 0 | 96 | 9 | 9 | 38 | 0.00019 | 0.00203 |
| 0 | 0.11048 | 0 | 930 | 7 | 3 | 53 | 0.00232 | 0.00165 |
| 0 | 0.30553 | 1 | 178 | 6 | 9 | 40 | 0.00833 | 0.0429 |
| 1 | 0.47066 | 1 | 702 | 9 | 5 | 25 | 0.00718 | 0.00424 |
| 0 | 0.21556 | 1 | 281 | 8 | 8 | 58 | 0.00746 | 0.04289 |
| 1 | 0.58326 | 0 | 771 | 9 | 3 | 32 | 0.00812 | 0.00438 |
| 1 | 0.5265 | 1 | 872 | 9 | 1 | 85 | 0.00121 | 0.00155 |
| 1 | 0.10168 | 1 | 431 | 8 | 7 | 98 | 0.00761 | 0.04326 |
| 0 | 0.10068 | 1 | 501 | 8 | 1 | 76 | 0.00347 | 0.00171 |
| 0 | 0.13576 | 0 | 570 | 6 | 4 | 43 | 0.00907 | 0.00398 |
| 1 | 0.17261 | 1 | 495 | 9 | 3 | 30 | 0.00247 | 0.00231 |
| 1 | 0.13682 | 1 | 95 | 6 | 1 | 61 | 0.00567 | 0.0066 |
| 1 | 0.31584 | 0 | 240 | 5 | 9 | 59 | 0.00092 | 0.00204 |
| 0 | 0.43588 | 0 | 196 | 7 | 3 | 61 | 0.0034 | 0.00266 |
| 0 | 0.16804 | 0 | 700 | 8 | 1 | 45 | 0.00176 | 0.00176 |
| 1 | 0.12694 | 1 | 799 | 9 | 8 | 59 | 0.00932 | 0.04284 |
| 0 | 0.27407 | 1 | 257 | 9 | 7 | 86 | 0.00721 | 0.04288 |
| 0 | 0.30945 | 0 | 793 | 9 | 6 | 82 | 0.00651 | 0.04286 |
| 0 | 0.54928 | 0 | 757 | 7 | 8 | 72 | 0.00638 | 0.04278 |
| 0 | 0.5133 | 1 | 967 | 10 | 8 | 37 | 0.00266 | 0.00196 |
| 1 | 0.23179 | 0 | 741 | 7 | 6 | 42 | 0.00521 | 0.00229 |
| 0 | 0.56907 | 1 | 148 | 9 | 9 | 52 | 0.00907 | 0.04301 |
| 0 | 0.32939 | 0 | 469 | 6 | 3 | 51 | 0.00985 | 0.00571 |
| 0 | 0.19069 | 1 | 551 | 9 | 9 | 36 | 0.00863 | 0.04275 |
| 1 | 0.56247 | 1 | 144 | 9 | 4 | 86 | 0.00651 | 0.04296 |
| 1 | 0.22357 | 0 | 555 | 7 | 9 | 69 | 0.00759 | 0.04286 |
| 1 | 0.45581 | 1 | 398 | 7 | 7 | 74 | 0.00612 | 0.01494 |
| 0 | 0.1288 | 0 | 443 | 5 | 8 | 39 | 0.002 | 0.00177 |
| 1 | 0.24925 | 1 | 948 | 9 | 2 | 77 | 0.00802 | 0.00745 |
| 1 | 0.52192 | 1 | 330 | 6 | 4 | 25 | 0.00339 | 0.00952 |
| 0 | 0.21738 | 0 | 486 | 6 | 2 | 27 | 0.00894 | 0.00367 |
| 0 | 0.29275 | 1 | 752 | 8 | 6 | 75 | 0.00603 | 0.0062 |
| 1 | 0.19253 | 0 | 567 | 8 | 3 | 69 | 0.00527 | 0.00294 |
| 0 | 0.24518 | 1 | 901 | 6 | 1 | 81 | 0.00906 | 0.00264 |
| 1 | 0.48899 | 1 | 177 | 8 | 3 | 26 | 0.00637 | 0.00543 |
| 0 | 0.57374 | 1 | 636 | 5 | 6 | 38 | 0.00793 | 0.00299 |
| 0 | 0.54314 | 1 | 176 | 7 | 2 | 84 | 0.00595 | 0.00748 |
| 1 | 0.3329 | 1 | 836 | 9 | 6 | 50 | 0.00489 | 0.00278 |
| 0 | 0.44107 | 1 | 156 | 10 | 0 | 73 | 0.00204 | 0.00213 |
| 0 | 0.26473 | 1 | 947 | 7 | 9 | 37 | 0.00934 | 0.00334 |
| 1 | 0.32647 | 1 | 652 | 7 | 5 | 44 | 0.00447 | 0.00242 |
| 1 | 0.51869 | 1 | 975 | 9 | 8 | 94 | 0.0054 | 0.01236 |
| 1 | 0.56957 | 1 | 702 | 10 | 7 | 24 | 0.00345 | 0.00872 |
| 0 | 0.21669 | 0 | 825 | 5 | 8 | 23 | 0.00085 | 0.00296 |
| 1 | 0.28796 | 0 | 461 | 8 | 8 | 39 | 0.00478 | 0.00411 |
| 0 | 0.26901 | 0 | 988 | 7 | 7 | 51 | 0.00884 | 0.04285 |
| 0 | 0.50117 | 1 | 607 | 7 | 1 | 66 | 0.0014 | 0.00154 |
| 1 | 0.53227 | 1 | 803 | 10 | 9 | 37 | 0.00222 | 0.00315 |
| 1 | 0.29791 | 1 | 1000 | 10 | 1 | 86 | 0.00065 | 0.0013 |
| 0 | 0.43199 | 0 | 1010 | 7 | 7 | 53 | 0.00582 | 0.04298 |
| 1 | 0.25172 | 0 | 105 | 9 | 5 | 90 | 0.00729 | 0.0442 |
| 1 | 0.54124 | 1 | 358 | 7 | 0 | 62 | 0.00365 | 0.00245 |
| 1 | 0.55773 | 0 | 1006 | 6 | 8 | 30 | 0.00452 | 0.00722 |
| 1 | 0.59994 | 0 | 211 | 6 | 8 | 50 | 0.00359 | 0.0133 |
| 1 | 0.21861 | 1 | 320 | 9 | 4 | 49 | 0.00783 | 0.04301 |

## Parameter normalization procedure for mathematical equation

All parameters in the mathematical equation that are developed in section 3.4. of the manuscript are normalized to the range [0, 1] using minimum and maximum values reported in Supplementary Table S3.

**Table S3**: The minimum and maximum values of all features used in the mathematical equation

| Parameter | Minimum | Maximum |
|---|---|---|
| $\delta$ | 1.02217 | 12.52157 |
| $T$ | 850 | 1891 |
| $\eta$ | -15.73156 | -0.05447 |
| $\sigma_{VEC}$ | 0.45826 | 3.15079 |
| $\sigma_{\Delta H_{\text{mix}}}$ | 0.96984 | 15.00423 |
| $\Phi$ | 0.16655 | 6.53409 |
| $\Delta H_{\text{mix}}$ | -51.4852 | 2.772 |
| $\Delta\chi$ | 1.4845 | 1.798 |
| $R_{\text{Mir}}$ | 128.05 | 147.92 |
| $k_1^{\mathrm{cr}}$ | -74.81401 | 44.62762 |
| $VEC$ | 3.65 | 7.8 |
| $\Omega$ | 0.28315 | 834.82288 |
| $Grp$ | 4.3 | 10.4 |
| $\Delta S_{\mathrm{mix}}$ | 9.11321 | 13.37745 |

## Supplementary references